\documentstyle[prd,aps]{revtex}

\begin{document}

\title{
Stability of quantum states of
 finite macroscopic systems
\footnote{
Proceedings of the Japan-Italy Joint Waseda Workshop on 
{\it Fundamental Problems in Quantum Mechanics}, 27-29 September, 2001, Tokyo, Japan. (Edited by S.\ Tasaki, To be published from World Scientific, 2002) }}

\author{Akira Shimizu, Takayuki Miyadera\cite{MIYA}
and Akihisa Ukena}

\address{
Department of Basic Science, University of Tokyo\\ 
3-8-1 Komaba, Tokyo 153-8902, Japan\\
E-mail: shmz@ASone.c.u-tokyo.ac.jp}


\maketitle

\begin{abstract}
We study the
stabilities of quantum states of macroscopic systems, 
against noises, against perturbations from environments, 
and against local measurements.
We show that the stabilities are closely related to 
the cluster property, which describes the strength of 
spatial correlations of fluctuations of local observables, 
and to fluctuations of additive operators.
The present theory has many applications, among which 
we discuss the mechanism of phase transitions in finite systems and 
quantum computers with a huge number of qubits.
\end{abstract}

\section{Introduction}\label{intro}

The stability of quantum states of macroscopic systems, 
which are subject to noises or perturbations from environments, have been 
studied in many fields of physics, 
including studies of 
`macroscopic quantum coherence' \cite{AL} and
quantum measurement.\cite{Zurek}
However, most previous works assumed
that the principal systems were describable 
by a {\it small number} of collective coordinates.
Although such models might be applicable to systems which 
have a non-negligible 
energy gap to excite `internal coordinates' of 
the collective coordinates, 
there are many systems that do not have such an energy gap.
As a result of the use of such simple models, 
{\em the results depended strongly on the choices of 
the coordinates and the form of the interaction} $\hat H_{\rm int}$
between the principal system and a noise or an environment.
For example, a robust state for some $\hat H_{\rm int}$ can become
a fragile state for another $\hat H_{\rm int}$.
However, macroscopic physics and experiences indicate that 
a more universal result should be 
drawn for the stability of quantum states of macroscopic systems.
Moreover, the stability against measurements were not studied well.
Although one might conjecture that effects of measurements would be equivalent
to effects of noises or environments, the conjecture is wrong as we will 
show in section \ref{sec-meas} of this paper.

In this paper, we study these stabilities using 
a general model with a {\em macroscopic} number of degrees of 
freedom $N$.
In addition to the fact that $N$ is huge, we make 
full use of the {\em locality} ---
`additive' observables must be 
the sum of local observables over a macroscopic region
(Eq.\ (\ref{A}) below), 
the interaction $\hat H_{\rm int}$ must be local (Eqs.\ (\ref{int-noise}) and 
(\ref{int-env})),
and measurement must be local.
By noticing these points, 
which were absent or ambiguous in most previous works,
we derive general and universal results.\cite{detail}
We also propose a new criterion of stability; the stability against  
local measurements.
We present a general and universal result also 
for this stability.\cite{detail}

The present theory has many applications
because it is general and universal.
We here mention applications to phase transitions in finite systems,
and quantum computers with a huge number of qubits.

\section{Macroscopic quantum systems}\label{sec-macro}

We consider ``macroscopic quantum systems.''
We first describe what this means.

As usual, we are only interested in phenomena in some energy 
range $\Delta E$,
and describe the system 
by an effective theory which correctly describes the system only 
in $\Delta E$.
For a given $\Delta E$, 
let ${\cal M}$ be the number of {\em many-body} quantum states in that 
energy range.
Then, 
\begin{equation}
N \sim \ln {\cal M}
\label{NM}\end{equation}
is the degrees of freedom of the effective theory.
Here, the symbol `$\sim$' means that 
corrections of $o(N)$, such as $\ln(\Delta E) = O(\ln N)$,
are neglected. 
For weakly interacting systems, for example, 
$N$ becomes the number of {\em single-body} quantum states 
which constitutes the ${\cal M}$ many-body states.
Note that $N$
sometimes becomes a small number 
even for a system of many degrees of freedom
when, e.g., a non-negligible energy gap exists in $\Delta E$.
Such systems include
some SQUID systems at low temperatures, 
and a heavy atom at a meV or lower energy range.
{\em We here exclude such systems}, 
because they are essentially systems of small degrees of freedom.
Namely, we consider 
systems whose $N$ is a macroscopic number.
Otherwise,  
the difference between $O(N)$ and $O(N^2)$, 
which plays the central roles in macroscopic physics and 
in the following discussions,  
would be irrelevant. 
Since $\Delta E$ sets 
a minimum length scale $\ell$, 
the system extends spatially over a finite volume
\begin{equation}
V \sim N \ell^d
\label{VvN}\end{equation}
in $d$ dimension.
Since $V$ is proportional to 
the macroscopic number $N$, we say $V$ is also macroscopic, 
disregarding the magnitude of $\ell$.

In short, we consider macroscopic quantum systems
for which $N$ and $V$ are macroscopic
for a given energy range $\Delta E$.
Note that $V$ is essentially equal to $N$
because of Eq.\ (\ref{VvN}). 
We therefore use $V$ and $N$
(and the words ``volume'' and ``degrees of freedom'')
interchangeably
in the following discussions.

\section{Cluster property}\label{sec-cluster}

As we shall show later, 
correlations between distant points are important in the study of stability.
As a measure of the correlations, we first consider the 
cluster property. 
Although we are considering finite systems, we first
review the case of infinite systems.\cite{haag}
A quantum state of an in {\em infinite} system 
 is said to have the {\it cluster property}
if 
\begin{equation}
\langle \delta \hat a(x) \delta \hat b(y) \rangle
\to 0
\mbox{ as } |x - y| \to \infty
\end{equation}
for any local operators $\hat a(x)$ and $\hat b(y)$
at $x$ and $y$, respectively, 
where 
\begin{eqnarray}
\delta \hat a(x) &\equiv& \hat a(x) - \langle \hat a(x) \rangle,
\\
\delta \hat b(y) &\equiv& \hat b(y) - \langle \hat b(y) \rangle.
\end{eqnarray}
Here, by a local operator at $x$ 
we mean a finite-order polynomial of 
field operators and their finite-order derivatives at position $x$.\footnote{
To express and utilize the locality of the theory manifestly, 
we use a field theory throughout this paper.}
The cluster property should not be confused with
the lack of long-range order: A state with a long-range
order can have the cluster property.
In fact, symmetry-breaking vacua of infinite systems
have {\em both} the long-range order and the cluster 
property.\cite{haag,ruelle,HL,miyashita,KT,pre01}
Many theorems have been proved regarding the cluster property.
\cite{haag,ruelle}
Among them, the following is most important to the present 
theory:

{\em 
Known theorem:
In infinite systems, 
any pure state has the cluster property. 
}

In oder to study finite systems,
we generalize the concept of the cluster property to
the case of finite systems.
For a small positive number $\epsilon$, 
let $\Omega(\epsilon)$ be the minimum size of the region
over which correlations of any local operators become smaller than $\epsilon$.
Namely, $\Omega(\epsilon) \equiv \sup_x |\Omega(\epsilon,x)|$, 
where $|\Omega(\epsilon,x)|$ denotes the size of the region 
$\Omega(\epsilon,x)$. Here, $\Omega(\epsilon,x)$ is defined by its complement 
$\Omega(\epsilon,x)^c$, which is the region of $y$ in which 
\begin{equation}
\left|
\langle \delta \hat a(x) \delta \hat b(y) \rangle
\right|
\leq
\epsilon
\sqrt{
\langle \delta \hat a^\dagger(x) \delta \hat a(x) \rangle
\langle \delta \hat b^\dagger(y) \delta \hat b(y) \rangle
}
\end{equation}
for any local operators $\hat a(x)$ and $\hat b(y)$.
Then, we say that a quantum state of a finite macroscopic system
has the {\it cluster property} 
if $\Omega(\epsilon) \ll V$
for sufficiently large $V$.
More strictly, we consider a sequence of states
for various values of $V$ such that 
they are essentially equivalent to each other except for the 
values of $V$.\footnote{
For example, the ground-state wavefunctions of many particles
for the same particle density, for various values of $V$.
} 
We say that the states (for large $V$) of the sequence
have the {\em cluster property} 
if $\Omega(\epsilon)$ for any $\epsilon > 0$ becomes
independent of $V$ for sufficiently large $V$.
It is clear from the known theorem above that 
a sequence of {\em pure} 
states of finite $V$ that do not have the cluster property
approaches a {\em mixed} state
of an infinite system as $V \to \infty$.

\section{Anomalously-fluctuating states}\label{sec-AFS}

As a second measure of correlations between distant points, 
we consider fluctuations of additive operators.
Here, by an {\em additive operator} we mean an
operator of the following form:
\begin{equation}
\hat A
=
\sum_{x \in V} \hat a(x),
\label{A}\end{equation}
where $\hat a(x)$ denotes a {\it local operator at $x$}.
When we regard the system as a composite system of subsystems 1 and 2, 
then
$\hat A = \hat A^{(1)} + \hat A^{(2)}$, 
hence the name ``additive.''
In thermodynamics, it is assumed that 
\begin{equation}
\langle \delta A^2 \rangle \leq o(V^2)
\end{equation}
for all additive quantities.\footnote{
Otherwise, it would be meaningless to talk about the average value
of $A$, which is $O(V)$.
}
In particular, if a state (classical or quantal) satisfies
\begin{equation}
\langle \delta A^2 \rangle \leq O(V)
\end{equation}
for all additive quantities, we call it a 
{\em normally-fluctuating state} (NFS).
In quantum theory of {\em finite} macroscopic systems, 
however, there exist pure states for which {\em some} of
additive operators have anomalously-large fluctuations;
\begin{equation}
\langle \delta \hat A^2 \rangle
= O(V^2).
\end{equation}
We call such a {\em pure} state an {\em anomalously-fluctuating state} (AFS).
\footnote{
It is possible to change an AFS into an NFS by enlarging the system
by adding an extra system of volume $V_{\rm extra} \sim V^2$, in which 
the quantum state is an NFS, because this leads to 
$\langle \delta A^2 \rangle = O(V^2)+O(V_{\rm extra})
=O(V_{\rm extra})
\simeq O(V_{\rm extra}+V)$.
We here exclude such an artificial and uninteresting case.
}

It is easy to show that 
{\em 
an AFS does not have the cluster property}. 
Hence, according to the known theorem of section \ref{sec-cluster}, 
an AFS cannot be a pure state in {\em infinite} systems.
Considering also that in thermodynamics any state in a pure phase is 
an NFS, we see that 
an AFS can exist only in {\em finite} macroscopic {\em quantum} systems.
Since AFSs are such unusual states, they might be expected to be
unstable in some sense.
Our purpose is to study this conjecture, by formulating 
the stability definitely, and thereby present general theorems
about the stability. 

\section{Fragility of quantum states of macroscopic systems}
\label{sec-fragile}

We say a quantum state is {\em fragile} if
its decoherence rate $\Gamma$ satisfies
\begin{equation}
\Gamma
\sim
K V^{1+\delta},
\label{r}\end{equation}
where $K$ is a function of microscopic parameters, 
and $\delta$ is a positive constant.

To understand the meaning of the fragility, 
consider first the non-fragile case where $\delta=0$.
In this case, the decoherence rate {\em per unit volume}
is independent of $V$. 
This is a normal situation in the sense that 
the total decoherence rate $\Gamma$ is 
basically the sum of {\em local} decoherence rates,
which are determined only by 
microscopic parameters (i.e., independent of $V$).
On the other hand, 
the case $\delta > 0$ is an anomalous situation in which 
the decoherence rate {\em per unit volume}
behaves as 
$
\sim
K V^{\delta}$.
Note that this can be very large even when $K$ is small,
because $V$ is huge.
This means that
{\it a fragile quantum state of a macroscopic system 
decoheres due to a noise or environment
at an anomalously great rate}, 
even when the coupling constant between 
the system and the noise or environment is small.

\section{Fragility in weak noises}
\label{sec-noise}

The most important assumption of the present theory is the locality.
For the interaction with a noise, the locality requires that the 
interaction Hamiltonian should be the sum of local interactions;
\begin{equation}
\hat H_{\rm int}
=
\sum_{x \in V}
f(x,t) \hat a(x).
\label{int-noise}\end{equation}
Here, 
$f(x,t)$ is a random noise field with vanishing average
$\overline{f(x,t)} = 0$, 
and $\hat a(x)$ is a local operator at $x$.
(See section \ref{sec-cluster} for the meaning of the local operator.)
We assume that the statistics of $f(x,t)$ is translationally invariant 
both spatially and temporally, i.e., 
$\overline{f(x,t) f(x',t')}$ is a function of 
$x-x'$ and $t-t'$.
We also assume that the time correlation of the noise is short.

The total Hamiltonian is 
\begin{equation}
\hat H_{\rm total}
=
\hat H + \hat H_{\rm int}.
\end{equation}
Here, $\hat H$ denotes the Hamiltonian of the 
principal system, which can be 
a general Hamiltonian including, e.g., many-body interactions.
Using this general local model,
we can show the following
for the fragility that is defined in section \ref{sec-fragile}:

\begin{em}
Theorem 1:
Let $|\Psi \rangle$ be a pure state,
whose time evolution by $\hat H$ is slow,
 of a macroscopic system.
If $|\Psi \rangle$ is an AFS, then 
it is fragile in the presence of some weak noise.
If $|\Psi \rangle$ is an NFS, then 
it is not fragile in any weak noise.
\end{em}

It follows from this theorem that 
an AFS decoheres (hence collapses) at an anomalously great rate
if external noises contain such a noise component, 
whereas a NFS does not decohere at such an anomalously great rate
in any weak noise.

\section{Fragility under weak perturbations from environments}
\label{sec-env}

The physical realities of noises are 
perturbations from environments.
We can show a similar theorem for
the effects of perturbations from environments.
Again, 
the most important assumption is the locality of
the interaction between the principal system and an environment.
Namely, the interaction Hamiltonian should be the sum of local interactions;
\begin{equation}
\hat H_{\rm int}
=
\sum_{x \in V}
\hat f(x) \hat a(x).
\label{int-env}\end{equation}
Here, $\hat f(x)$ and $\hat a(x)$ are local operators at $x$ of
an environment and the principal system, respectively. 
Similarly to the case of noise, 
we assume that (in the interaction picture)
$\langle \hat f(x,t) \rangle_{\rm E} = 0$, 
and that
$\langle f(x,t) f(x',t') \rangle_{\rm E}$ is a function of 
$x-x'$ and $t-t'$, 
where $\langle \cdots \rangle_{\rm E}$ denotes
the expectation value for the state of the environment E.
We also assume that the correlation time 
of $\langle f(x,t) f(x',t') \rangle_{\rm E}$ is short.
The total Hamiltonian is 
\begin{equation}
\hat H_{\rm total}
=
\hat H + \hat H_{\rm int} + \hat H_{\rm E},
\end{equation}
where $\hat H$ and $\hat H_{\rm E}$ denote the Hamiltonians of the 
principal system and the environment, respectively.
Here, $\hat H$ can be 
a general Hamiltonian including, e.g., many-body interactions.
Using this general local model,
we can show the following
for the fragility that is defined in section \ref{sec-fragile}:

\begin{em}
Theorem 2:
Let $|\Psi \rangle$ be a pure state,
whose time evolution by $\hat H$ is slow,
of a macroscopic system.
If $|\Psi \rangle$ is an AFS, then 
it is fragile under some weak perturbation from some environment.
If $|\Psi \rangle$ is an NFS, then 
it is not fragile under any weak perturbations from environments.
\end{em}

It follows from this theorem that
an AFS decoheres (hence collapses) at an anomalously great rate
if perturbations from environments contain such a perturbation term, 
whereas a NFS does not decohere at such an anomalously great rate
under any weak perturbations from environments.

\section{Do relevant perturbations always exist?}\label{sec-exist}

By theorems 1 and 2, 
we have shown that NFSs are not fragile
in any noises or environments, which interact weakly with
the principal system via any local interactions.
This should be contrasted with the results of most previous works, 
according to which a state could be either fragile or robust 
depending on the form of the interaction.
We have obtained the general and universal conclusion 
because we have made full use of the locality as well as 
the huge degrees of freedom.

Regarding AFSs, on the other hand, theorems 1 and 2 show only that
they are fragile in {\em some} noise or environment,
which interact weakly with
the principal system via local interactions.
In other words, 
for any AFS
it is always possible to {\em construct} 
a noise (or an environment) and a weak local interaction that 
make the AFS fragile.
These theorems do {\em not} guarantee the {\em existence} 
of such 
a relevant noise (or an environment) and a relevant interaction
{\em in real physical systems}.
We discuss this point in this section.

As described in section \ref{sec-macro}, 
we are only interested in phenomena in some energy 
range $\Delta E$, and describe the system 
by an effective theory which correctly describes the system only 
in $\Delta E$.
The effective theory can be constructed from 
an elementary dynamics by an appropriate renormalization 
process.
In this process, in general, 
many interaction terms would be generated in 
the effective interaction $H_{\rm int}$.\footnote{
We expect, in accordance with experiences, 
that by an appropriate renormalization process 
$H_{\rm tot}$ can be made local in the relevant space-time
scale.
}
Hence, it seems quite rare that 
a relevant noise or an environment and 
a relevant interaction are {\em completely} absent.
Even when the coupling constant to
the relevant noise (or environment) is, say,  
ten times smaller than those to other noises (or environments), 
the relevant noise (or environment) 
would dominate the decoherence process of the AFS
because the decoherence rate grows anomalously fast 
with increasing $V$, 
except when the noise (or the perturbation from the environment)
is negligibly weak
such that its intensity is, e.g., $O(1/V)$.
Namely, 
an AFS should be fragile apart from such an exceptional case.

However, for general systems, 
we cannot exclude the exceptional case where 
the relevant noise is negligibly weak.
Therefore, 
we cannot draw a definite 
conclusion on whether AFSs are {\em always} fragile in 
{\em real} physical systems.
This motivates us to explore another stability, 
which will be described in the next section.

\section{Stability against local measurements}
\label{sec-meas}

We can prove a stronger statement by considering 
the stability against measurement.

Suppose that one performs an ideal measurement of a local 
observable $\hat a(x)$  at $t=t_a$
for a state $\hat \rho$ (pure or mixed) of a macroscopic system, 
and obtains a value $a$.
Subsequently, one measures another local 
observable $\hat b(y)$ at a later time $t_b$,
and obtains a value $b$.
Let $P(b;a)$ be the probability distribution of $b$, 
i.e., the probability that $b$ is obtained at $t_b$ under the condition
that $a$ was obtained at $t_a$.
On the other hand, 
one can measure $\hat b(y)$ at $t=t_b$ without performing 
the measurement of $\hat a(x)$ at $t_a$.
Let $P(b)$ be the probability distribution of $b$ in this case.
We say $\hat \rho$ is {\it stable against local measurements} if
for any $\epsilon >0$
\begin{equation}
\left| P(b;a) - P(b) \right| \leq \epsilon
\
\mbox{for sufficiently large $|x-y|$,}
\end{equation}
for {\em any} local operators $\hat a(x)$ and $\hat b(y)$
and their eigenvalues $a$ and $b$ such that 
$P(a) \geq \varepsilon$.

This stability is stronger than the stability against noises and 
perturbations from environments.
In fact, the latter stability is related to 
$| \sum_a P(b;a) - P(b) |$.
There are many examples of states for which 
$| \sum_a P(b;a) - P(b) | \leq \epsilon$ is satisfied whereas
$| P(b;a) - P(b) | \leq \epsilon$ is not.

For the simple case
 $t_b - t_a \to 0$, we can show the following:\footnote{
Results for more general cases will be described elsewhere.
}

\begin{em}
Theorem 3:
Let $\hat \rho$ be a pure or mixed state of a macroscopic system.
If $\hat \rho$ is stable against local measurements, then 
it has the cluster property, and vice versa.
\end{em}

It follows from this theorem that 
any AFS is {\em unstable} against local measurements.

\section{Mechanism of symmetry breaking in finite systems}
\label{sec-SB}

AFSs generally appear in, e.g., 
finite systems which will exhibit symmetry breaking if 
$V$ goes to infinity.
In such systems, 
we can find states (of finite systems)
which approach a symmetry-breaking vacuum as $V \to \infty$.
We call such a state a pure-phase vacuum.
It has a finite expectation value $\langle \hat M \rangle = O(V)$ of 
an additive order parameter $\hat M$, and 
has relatively small fluctuations 
$\langle \delta \hat A^2 \rangle \leq O(V)$
for {\em any} additive operator $\hat A$ (including $\hat M$)
\cite{ruelle,HL,miyashita,KT,pre01}.
Hence, the pure-phase vacua are NFSs.
In a mean-field approximation, 
the pure-phase vacua have the lowest energy.
However, 
it is {\em always} possible 
to construct a pure state(s) 
that does not
break the symmetry, $\langle \hat M \rangle = 0$, and has
an equal or {\em lower} energy than the pure-phase vacua \cite{HL,pre01}.
Although such states cannot be pure in infinite systems, 
they can be pure in finite systems \cite{haag,HL,pre01,jpsj02}.
When $[\hat H, \hat M] \neq 0$, in particular, 
the exact lowest-energy state 
is generally such a symmetric ground state\cite{HL,miyashita,KT,pre01}.
To lower the energy of a pure-phase vacuum, a SB field is necessary.
However, an appropriate SB field 
would not always exist in laboratories.
The symmetric ground state is composed primarily of a superposition of
pure-phase vacua with different values of $\langle \hat M \rangle$, 
and, consequently, 
it has an anomalously large 
fluctuation of $\hat M$; 
$\langle \delta \hat M^2 \rangle = O(V^2)$.\cite{HL,miyashita,KT,pre01}
Therefore, if one obtains 
the exact lowest-energy state 
(e.g., by numerical diagonalization) 
of a finite system, which will exhibit symmetry breaking if 
$V$ goes to infinity, the state is often
an AFS.

The present results suggest a new origin of symmetry 
breaking in finite systems.\cite{prl00}
Although symmetry breaking is usually described as a property of
infinite systems, it is observed in finite systems as well.
The results of sections \ref{sec-noise} and \ref{sec-env} suggest that 
although a pure-phase vacuum (which is an NFS) has
a {\em higher} energy than the symmetric ground state (an AFS),
the former would be realized because the latter is fragile
in some noises or environments.
This mechanism may be called ``environment-induced symmetry breaking,''
a special case of which was discussed
for interacting many-bosons.\cite{pre01,jpsj02,prl00}
The result of section \ref{sec-meas} 
suggests more strongly 
that only a pure-phase vacuum should be realized, because
an AFS is changed into another state
when one measures only a tiny part of the system, 
and such drastic 
changes continue by repeating measurements, until
the state becomes an NFS.
This mechanism may be called ``measurement-induced symmetry breaking.''

We consider that 
these scenarios
explain the symmetry breaking (i.e., realization of an pure-phase vacuum)
in finite systems, much more 
naturally and generally than the idea of the symmetry breaking field:
It seems quite artificial to assume that 
an appropriate static symmetry breaking field would always present 
in real physical systems,\footnote{
For example, it is quite unlikely that 
a SB field for antiferromagnets could exist in laboratories.
} 
 although it is true that
symmetry breaking fields are a convenient mathematical tool.

\section{Stability of quantum computers with many qubits}
\label{sec-QC}

Quantum computers are useful only when the number of qubits $N$ is huge.
Hence, useful quantum computers are macroscopic quantum systems.

Various states appear in the course of 
a quantum computation.
Some state may be an NFS, for which
$\langle \delta \hat A^2 \rangle = O(V)$
for {\em any} additive operator $A$.
This means that correlations between distant qubits are 
weak.
Properties of such states may be possible to emulate by a classical 
system with local interactions.
We therefore conjecture that 
other states -- AFSs -- should
appear in some stages of the computation for a quantum computer to be 
much faster than classical computers.\cite{crest}
In fact, two of the authors 
confirmed this conjecture in Shor's algorithm for 
factoring.\cite{US}

The present results suggest that 
the decoherence rate of quantum computers can be estimated by fluctuations of 
additive operators, which depend strongly on the 
number of qubits $N$ and the natures of the states 
of the qubits.\cite{ekert}
Since AFSs are used in some stages of the fast quantum computation,
the state of qubits can become fragile in some noise or
environment, for quantum computers with many qubits.
Note that the dominant perturbation for the case of huge $N$ 
can be different from that for small $N$, 
because the decoherence rate of an AFS grows anomalously 
fast with increasing $N$.
Therefore, the quantum computer should be designed in such a way that
it utlizes AFSs for which the intensities of the relevant noises 
are $O(1/N)$ or smaller.
Since the error corrections are not almighty, 
we think that one must consider both such optimization and 
the error corrections to realize a quantum computer 
with a large number of qubits.

\section{Discussions}\label{sec-discuss}

The present results show that the stabilities of 
quantum states of finite macroscopic systems are
closely related to the cluster property, which describes the strength of 
spatial correlations of fluctuations of local observables, 
and to fluctuations of additive operators.
Note that the stabilities are defined as {\em dynamical} 
properties of an {\em open}
system, whereas the cluster property and fluctuations of additive operators
are defined as {\em static} properties of a {\em closed} system.
Hence, it is non-trivial --- may be 
surprising --- that they are closely related to each other.

We stress that the {\em approximate} stability
against {\em all} local interactions (between the principal system
and environments) would be more important than the {\em exact} stability
against a {\em particular} interaction, which was frequently discussed
in previous works.
As discussed in section \ref{sec-exist},
many types of interactions would coexist in real physical systems, 
and the exact stability against one of them could not exclude 
fragility to another.

In this paper, we did not mention temperature.
It is clear that similar conclusions can be drawn for
thermal equilibrium states 
(Kubo-Martin-Schwinger states\cite{haag,ruelle}), because
thermal equilibrium states can be represented as vector states
by introducing an auxiliary field.\cite{TFD}

We also point out that the present results may be important 
to study the foundations of 
non-equilibrium statistical physics.
For example, in the linear response theory of Kubo,\cite{kubo} 
he assumed the unitary time evolution of a closed system.
However, actually, the system is continuously measured 
over a time period longer than $1/\omega$
when one measures, say, the AC conductivity at frequency $\omega$.
Therefore, it is necessary 
for the validity of the linear response theory 
that the non-equilibrium state under consideration 
is stable against measurements.
Theorem 3 suggests that such states must have the cluster property. 
This observation 
may become a foundation not only of non-equilibrium statistical physics
but of non-equilibrium field theory, which is not established yet.

\section*{Acknowledgments}
The authors thank
Prof.\ I.\ Ojima for discussions and suggestions.

\end{document}